\documentclass[fleqn,twoside]{article}
\usepackage{espcrc2}
\usepackage{latexsym}

\usepackage{graphicx}
\usepackage[figuresright]{rotating}

\newcommand{\AmS}{{\protect\the\textfont2
  A\kern-.1667em\lower.5ex\hbox{M}\kern-.125emS}}

\title{Moving and staying together without a leader}

\author{Guillaume Gr\'egoire,\address[CEA]{CEA -- Service de Physique de l'Etat Condens\'e,
Centre d'Etudes de Saclay,\\
91191 Gif-sur-Yvette, France}
Hugues Chat\'e,\addressmark
and Yuhai Tu\address[IBM]{IBM T.J. Watson Research Center, Yorktown Heights, 
NY 10598, USA}}
\newcommand\lbt{\left<}
\newcommand\rbt{\right>}
\newcommand\lpa{\left(}
\newcommand\rpa{\right)}
\newcommand\lcr{\left[}
\newcommand\rcr{\right]}
\newcommand\bequ{\begin{equation}}
\newcommand\eequ{\end{equation}}
\newcommand\barr{\begin{eqnarray}}
\newcommand\earr{\end{eqnarray}}

\begin{document}

\begin{abstract}
A microscopic, stochastic, minimal model for collective and cohesive motion
of identical self-propelled particles is introduced. 
Even though the particles 
interact strictly locally in a very noisy manner, we show that cohesion
can be maintained, even in the zero-density limit of an arbitrarily
large flock in an infinite space. The phase diagram spanned by the
two main parameters of our model, which encode the tendencies for
particles to align and to stay together, contains non-moving ``gas'',
``liquid'' and ``solid'' phases separated from their moving counterparts
by the onset of collective motion.
The ``gas/liquid'' and ``liquid/solid'' are shown to be first-order 
phase transitions in all cases. In the cohesive phases, we study also
the diffusive properties of individuals and their relation to
the macroscopic motion and to the shape of the flock.
\end{abstract}

\maketitle

\section{Introduction}

The emergence of collective motion of self-propelled organisms 
(bird flocks, fish schools, herds, slime molds, bacteria colonies, etc.) 
is a fascinating phenomenon which attracted the attention of 
(theoretical) physicists only recently 
(\cite{Parrish,VICSEK,Shimoyama,Levine,Duparc}).
Particularly intriguing are the situations where no ``leader'' with 
specific properties is present in the group, no mediating
field helps organizing the collective dynamics (e.g. no chemotaxis),
and interactions are short-range. In this case, even the possibility of
collective motion may seem surprising.

However ``simple'' the involved organisms may be, they are still
tremendously complex for a physicist and his inclination will often be
to go away from the detailed, intricate, as-faithful-as-possible
modeling approach usually taken by biologists, and to adopt 
``minimal models'' hopefully catching the crucial, universal
properties which may underlie seemingly different situations.

In this setting, the organisms can be reduced to points
which move at finite velocity and interact  with 
neighbors. This is in fact what Vicsek and collaborators did when
introducing their minimal model for collective motion.

\section{Vicsek's model}

Vicsek's model \cite{VICSEK} consists in  pointwise particles
labeled by $i$
 which move synchronously at discrete timesteps $\Delta t$
 by a fixed distance $v_0$
along a direction $\theta_i$. This angle is
calculated from the current velocities
of all particles $j$ within an interaction range $r_{\rm 0}$,
reflecting the only ``force'' at play,
a tendency to align with neighboring particles:
\begin{equation}
\theta_i^{t+1} = \arg \left[
\sum_{j\sim i} \vec{v}_j^{t} \right] + \eta \, \xi _i^t  \;,
\label{corerule}
\end{equation}
where $\vec{v}_i^t$ is the velocity vector of magnitude $v_0$ along
direction $\theta_i$ and
$\xi _i^t$ is a delta-correlated white noise ($\xi\in [-\pi,\pi]$).
Fixing $r_{\rm 0} =1$, $\Delta t=1$, and choosing, without loss of
 generality, a value $v_0 \Delta t< r_{\rm 0}$, Vicsek {\it et al} studied 
the behavior  of this simple model in the  two dimensional parameter space 
formed by the noise strength
$\eta$ and $\rho$, the particle density. They found, at large $\rho$ and/or
small $\eta$, the existence of an ordered phase characterized by:
\begin{equation}
V \equiv {\left\langle | {\langle\vec{v}_i^{\/t}\rangle}_i | 
\right\rangle}_t > 0 \;,
\label{eqV}
\end{equation}
i.e. a domain of parameter space in which the particles move collectively.

The existence of the ordered phase was later proved 
analytically \cite{TT} {\it via}
a continuous model for the coarse-grained particle velocity and density.
Vicsek {\it et al} devoted most of their effort to studying the transition
to the ordered phase \cite{VICSEK}.
They found numerically a continuous transition
characterized by scaling laws and they tried to estimate the corresponding
set of critical exponents.

\section{Collective and cohesive motion}

Vicsek's model accounts rather well, at least at a qualitative level,
for situations where the organisms interact at short distances but
need not stay together. This is for instance the case of the
bacterial bath recently studied by Wu and Libchaber \cite{WL}.
\begin{table}
\begin{tabular}{|c|c|c|c|c|c|c|} \hline
$\Delta t$ & $v_0$ & $r_{\rm 0}$ & $r_{\rm a}$ & $r_{\rm e}$ & $r_{\rm c}$ & $\eta $ \\ \hline
1.0 &  0.05 & 1.0 & 0.8 & 0.5 & 0.2 & 1.0 \\ \hline
\end{tabular}
\caption{Fixed-value parameters used in the simulations.}\label{t1}
\end{table}
 In this experiment,
{\it E. Coli} bacteria are swimming freely within a fluid film of 
thickness approximately equal to their size. By seeding the system with
polystyrene beads and recording the trajectories of these passive tracers,
Wu and Libchaber showed that the bacteria perform superdiffusive motion 
crossing over to normal diffusion. We later argued that the superdiffusive
behavior is likely to be due to the onset of collective motion as in
Vicsek's model \cite{COMMENT,WL-REPLY}.

When the situation to be described involves the overall cohesion of the 
population, Vicsek's model needs to be supplemented by a suitable
feature. Indeed, an initially cohesive flock of particles will disperse 
in an open space.
In other words, no collective motion is possible in the zero density limit 
of this model. In the following, we extend Vicsek's model to account
for the possible cohesion of the population of particles.

The above remark is by no means new. Early models for the 
collective motion of  ``boids'' (contraction
of ``birdoid'', a term used by computer animation graphics specialists)
do include a two-body repulsive-attractive interaction \cite{Reynolds}.
More recent works by physicists also included this ingredient, but 
they either comprised an extra global interaction \cite{Shimoyama},
or the actual interaction range used extended over the whole flock
for the sizes considered, making it effectively global
\cite{Levine}. Another encountered pitfall, from our point of view at least,
is to enforce the cohesion by the confinement to a rather small, close,
space \cite{Duparc}. 

Here we want to be, in a sense, in the least-favorable circumstances
for observing collective motion: no leader in the group, strongly noisy
environment and/or communications, strictly local interactions, and
no confinement at all. The ``minimal'' model presented below is one
of the simplest possible ones satisfying these constraints.

\section{A minimal model}

In addition to the possibility of achieving cohesion, we also want
to confer a ``physical'' extent to the particles, a feature absent from
Vicsek's point-particles approach.
Adding a Lennard-Jones-type body force $\vec{f}$
acting between each
pair of particles within distance $r_{\rm 0}$ from each other
offers such a possibility.

Equation~(\ref{corerule}) is then replaced by
\begin{equation}
\theta_i^{t+1} = \arg \left[
\alpha\sum_{j\sim i} \vec{v}_j^t
+ \beta \sum_{j\sim i}  {\vec{f}}_{ij} \right]
+ \eta \, \xi _i^t   \;,
\label{boidrule1}
\end{equation}
where $\alpha$ and $\beta$ control the relative importance of the
two ``forces''.
The precise form of the dependence of the body force on the distance
between the two particles involved is not important. It is enough to
ensure a hard-core repulsion at distance $r_{\rm c}$ 
and an ``equilibrium'' preferred distance $r_{\rm e}$.

In the following, we use
\begin{equation}
{\vec{f}}_{ij} = \vec{e}_{ij}\left\{
\begin{array}{ll}
-\infty & {\rm if}\;\;r_{ij}<r_{\rm c}\;,\\
\frac{1}{4}\frac{r_{ij}-r_{\rm e}}{r_{\rm a}-r_{\rm e}}
 & {\rm if} \;\; r_{\rm c}<r_{ij}<r_{\rm a}\;,\\
1 & {\rm if}\;\; r_{\rm a}<r_{ij}<r_{\rm 0}\;.
\end{array} \right.
\label{bodyforce}
\end{equation}
with $r_{ij}$ the distance between boids $i$ and $j$, $\vec{e}_{ij}$
the unit vector along the segment going from $i$ to $j$, and the numerical
 values $r_{\rm c}=0.2$, $r_{\rm e}=0.5$ and $r_{\rm a}=0.8$.

\begin{figure*}
\begin{center}
\includegraphics*[width=14cm]{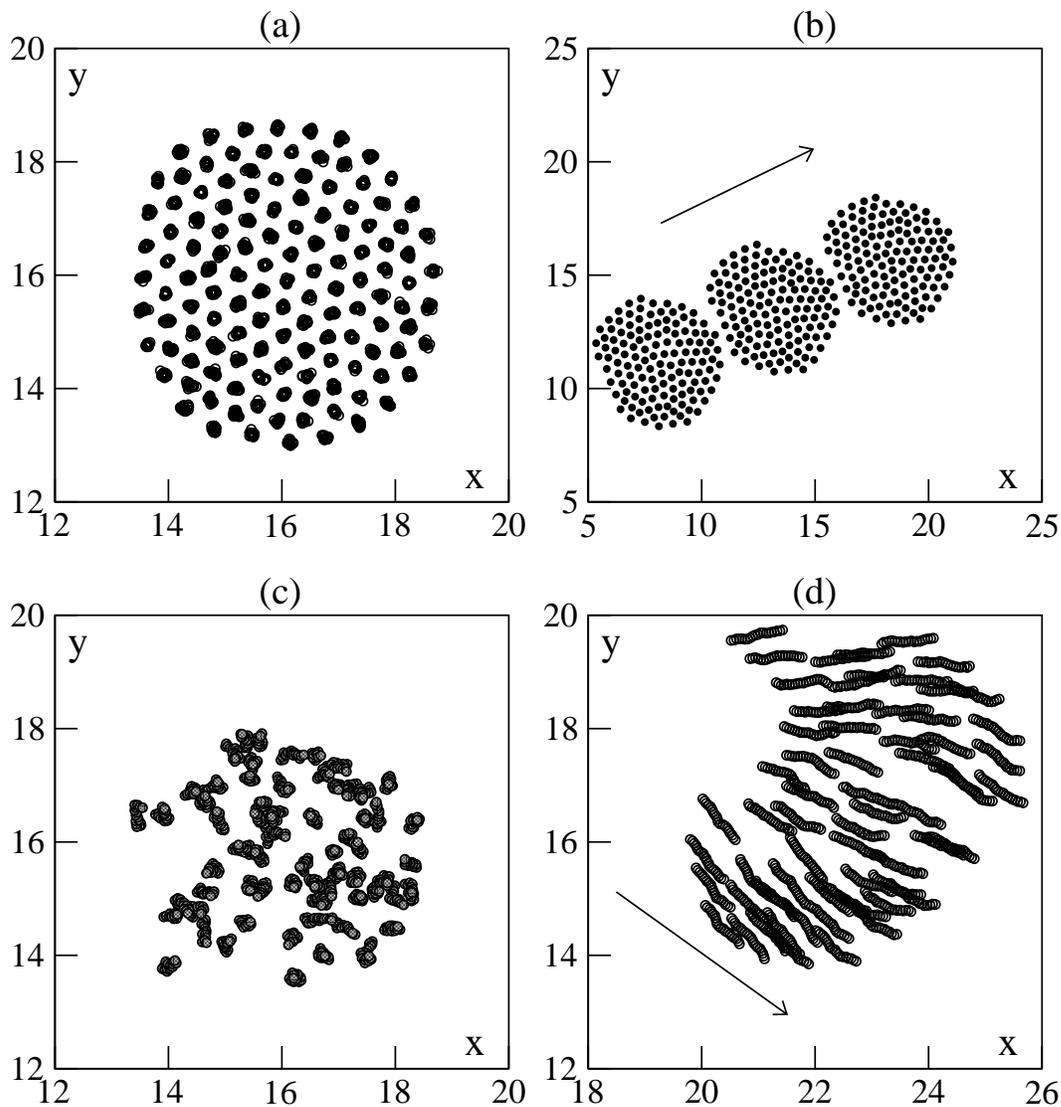}
\caption{Cohesive flocks of 128 particles in a square box of linear size
32 with periodic boundary conditions (for parameters see Table~\protect\ref{t1}).
(a): immobile ``solid'' at $\alpha=1.0$ and $\beta=100.0$ (20 timesteps
superimposed).
(b): 3 snapshots, separated by 120 timesteps, 
of a ``flying crystal'' at  $\alpha=3.0$ and $\beta=100.0.$
(c): fluid droplet ($\alpha=1.0$, $\beta=2.0$, 20 consecutive timesteps).
(d): moving droplet ($\alpha=3.0$, $\beta=3.0$, 20 consecutive timesteps).
In (b) and (d), the arrow indicates the (instantaneous) direction of motion.}
\label{fig01}
\end{center}
\end{figure*}

\begin{figure*}
\begin{center}
\includegraphics*[width=13cm]{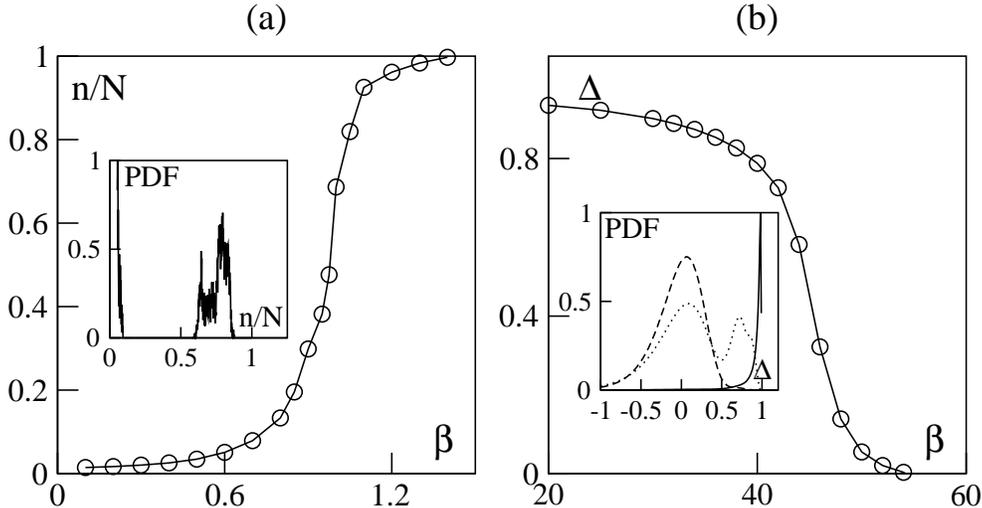}
\caption{Order parameters at $\rho =1/16$, $L=128$, $\alpha =1.0$. 
(a) : ``gas/liquid'' transition, inset: cluster mass 
distribution at the coexistence point $\beta =1.0$. 
(b) : ``liquid/solid'' transition, inset: pdf of order parameter 
$\Delta$ for the ``liquid/solid'' transition, 
$\beta =36$ plain line (``liquid'' phase), $\beta =48$ dotted line and 
$\beta =60$ dashed line(``solid'' phase).
 (other parameters as in Table~\protect\ref{t1}).
 }
\label{fig02}
\end{center}
\end{figure*}
We have also tested other types of noise term in the model.
In particular, considering the noise as the uncertainty with which each
boid ``evaluates'' the force exerted on itself by the $N_i$ neighboring boids 
leads to change Eq.~(\ref{boidrule1}) to:
\begin{equation}
\theta_i^{t+1} = \arg \left[
\alpha\sum_{j\sim i} \vec{v}_j^t
+ \beta \sum_{j\sim i}  {\vec{f}}_{ij} \;+\;N_i \eta \vec{u}_i^t \right]
\label{boidrule1bis}
\end{equation}
where $\vec{u}_i^t$ is a unit vector of random orientation.
There are delicate issues related to the choice of the noise term,
in particular with respect to the critical properties of the
transition to collective motion \cite{TBP}.
We mostly considered, in the following, the noise term
as prescripted above in Eq.~(\ref{boidrule1bis}). 
A detailed analysis of the influence of the 
nature of the noise is left for future studies, but we are confident that
the results presented here hold generally, at least at a qualitative level.

Finally, in order to ensure that each particle only interacts
with its ``first layer'' of neighbors (within
distance $r_{\rm 0}$), we calculate, at each timestep, the Voronoi tessellation
of the population~\cite{Voronoi}---this has the additional advantage of 
providing a natural definition of the ``cells'' associated with each
particle. The interacting neighbors are then retricted to be those
of the particles
within distance $r_{\rm 0}$ which are also neighbors in the Voronoi sense.

\section{Typical phases}

One can easily guess the ``phases'' that the above model can exhibit
for a fixed noise strength $\eta$ ($\eta=1.0$ in the following, 
for a summary of parameters see Table~\ref{t1}). 
We now present them in a qualitative manner. The results presented below were
all obtained in the two-dimensional case, but most of them hold in three
dimensions.
\begin{figure}
\begin{center}
\includegraphics*[width=6.5cm]{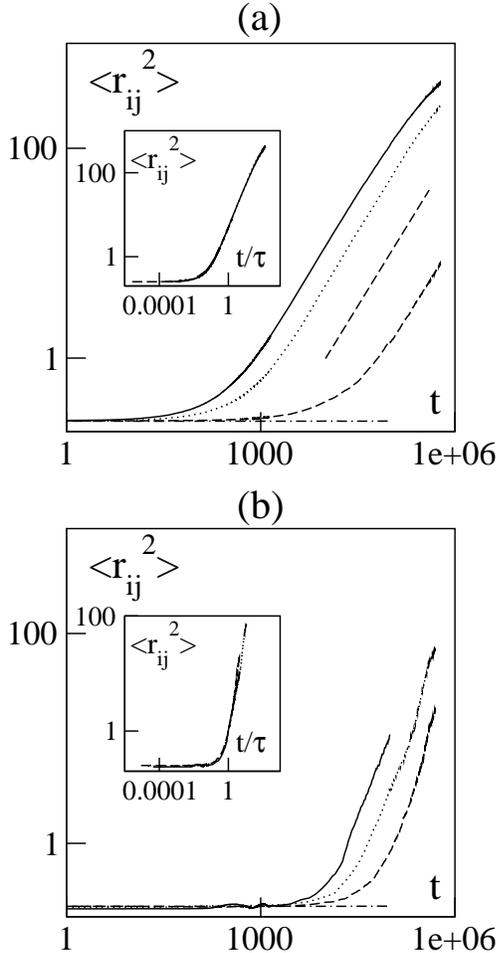}
\caption{Mean square distance between initially-neighboring particles
vs time for a flock of $10000$ boids ($L=400, \rho =1/16$) in
logarithmic scales.
(a) non-moving cohesive droplet ($\alpha =1.0$, $\beta =25.0$, $35.0$, $45.0$ 
and $150.0$ from top to bottom, the dashed line has slope 1); 
(b) moving cohesive flock ($\alpha =3.0$, $\beta =40.0$, $55.0$, 
$75.0$ and $150.0$ from top to bottom). 
Asymptotic transition points were measured at 
$\alpha =1.0$, $\beta_{\rm LS}=45.3$ and at $\alpha =3.0$, $\beta_{\rm LS}=76.1$ 
(see below). Other parameters as in Table~\protect\ref{t1}.
Insets: data collapse from which the waiting time $\tau$ can be estimated.}
\label{fig03}
\end{center}
\end{figure}
When the body force is weak
(small $\beta$ values), the cohesion of a flock cannot be maintained.
In a finite box (finite particle density $\rho$), 
one is left with a gas-like phase (not shown). 
An arbitrarily large flock in an infinite
space disintegrates, eventually leaving isolated random-walking particles.

For large enough $\beta$, we can expect the cohesion to be maintained.
Figure~\ref{fig01} shows such cohesive flocks.
The internal structure of the flocks depends also on $\beta$:
for large body force, positional quasi-order is present,
and the particles locally form an hexagonal crystal (Fig.~\ref{fig01}ab).
For intermediate values of $\beta$, no positional order arises,
and the flock behaves like a liquid droplet  (Fig.~\ref{fig01}cd).

The influence of the alignment ``force'' is manifested by the
global motion of the flock: for large enough $\alpha$ values,
the flock moves ($V>0$). Depending on $\beta$,
one has then either a ``moving droplet'' (Fig.~\ref{fig01}d) or a 
``flying crystal'' (Fig.~\ref{fig01}b).

\section{Order parameters}\label{Sorder}

Order parameters have to be defined to allow for a quantitative
distinction between the phases described above. 

The limit of cohesion
separating the ``liquid'' phases from the ``gas'' can be determined
by measuring the distribution of the sizes of particles clusters,
thanks to an implementation of the Hoshen-Kopelman \cite{HK} algorithm.
A cohesive flock is then one for which $n$, the size of the largest cluster,
is of order $N$, the total number of particles.
Below, we use the criterion $n/N = \frac{1}{2}$ to define the transition.
Increasing $\beta$,  $n/N$ sharply rises to order-one values
(Fig.~\ref{fig02}a).

\begin{figure}
\begin{center}
\includegraphics*[width=6.5cm]{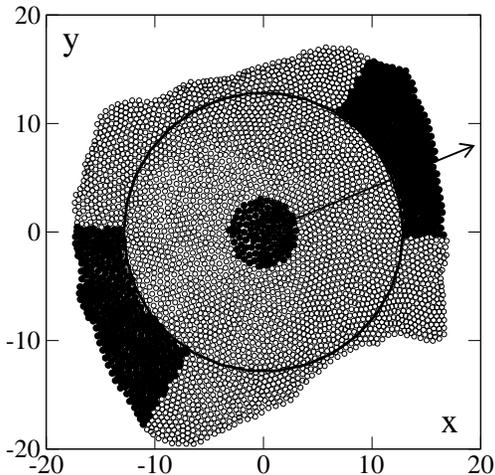}
\caption{Snapshot of a moving flock of 4096 boids (coordinates
centered on the position of the center of mass (CoM)). The arrow indicates the
instantaneous direction of motion of the flock. The solid circle,
centered on the CoM, has a radius equal to the root-mean-square
of all boids' distances to the CoM. The circle with a 4 times smaller
radius defines the  ``core'' region (filled circles near
the CoM). The ``head'' contains all boids outside the larger circle
which are also within a cone of opening angle $\pi/4$ centered on the
direction of motion. The ``tail'' region is defined in an opposite manner. 
}
\label{fig04}
\end{center}
\end{figure}

\begin{figure}
\begin{center}
\includegraphics*[width=6.5cm]{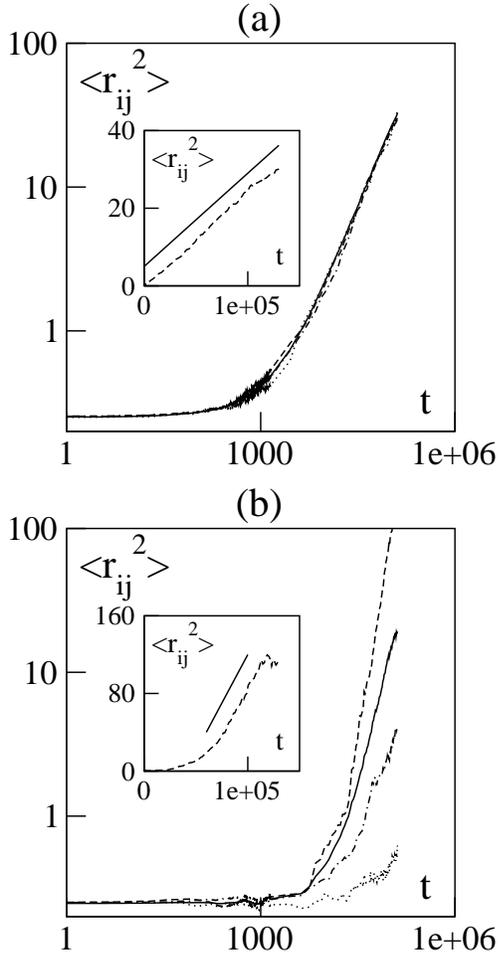}
\caption{Growth of the mean square distance between initially-neighboring 
particles depending on their position within a cohesive ``liquid'' flock
of $10000$ boids ($\rho =1/16$, $\beta=40$). (a) $\alpha =1.0$,
cohesive non-moving droplet; (b) $\alpha =3.0$, moving droplet.
Solid line: all boids in flock.
Dotted line: core region,
dashed line: head region,
dash-dotted line: tail region
 (as defined in Fig.~\protect\ref{fig04})
Insets : diffusion within flock head in linear scales, the 
solid line is just a guide to the eye.}
\label{fig05}
\end{center}
\end{figure}

The ``liquid/solid'' transition takes place when $\beta$ is large enough
so that cohesion of the population is ensured. 
To determine this onset of positional quasi-order within a finite
but arbitrarily large cohesive flock, 
we first observe that,
whether in the collective motion region or not, the ``liquid'' and
``solid'' phases can be distinguished by the fact that particles
diffuse with respect to each other in the ``liquid'', whereas
neighboring particles always remain close to each other in the ``solid''
(Fig.~\ref{fig03}).
To be more precise,  in the ``liquid'',
initially close-by particles remain so for some
trapping time $\tau$ (which can be defined or used
to collapse the curves of Fig.~\ref{fig03}). 
Approaching the ``solid'' phase
(by increasing $\beta$), $\tau$ diverges.
In addition, since we are not dealing with a translation-invariant
system, we should also distinguish between the diffusion properties
of the particles depending on their relative position within
the flock.
 We have thus defined different ``sectors'' (``core'', 
``head'', ``tail'', ``sides'') as explained in Fig.~\ref{fig04}.
If all sectors are roughly equivalent in non-moving droplets 
(at least sufficiently far from the ``liquid/solid'' transition), 
some differences are observed within moving droplets: 
the outer regions and in particular the head
are more active, whereas cohesion is stronger in the core (Fig.~\ref{fig05}).
Thus, in principle, different trapping times $\tau$ can be defined 
for the different regions of a (moving) flock. But the depth of the 
outer, ``more liquid'', layer does not depend on the flock size
(if it is big enough), so that, in large flocks, most of the population
behaves as the core. Consequently, the relative diffusion averaged over
the whole flock suffers from finite-size effects, but they disappear
in the large-size limit (see below). 
To sum up, the trapping time $\tau$ (measured on all particles of 
the flock) is a good quantity to track the ``liquid/solid'' transition.
However, instead of directly estimating of $\tau$, we measured
$\Delta$, the relative diffusion {\it over some large time $T$}
 of initially-neighboring  particles:
\begin{equation}
\Delta\equiv\left\langle \frac{1}{n_i}\sum_{j\sim i} 
\left( 1-\frac{r_{ij} ^2(t)}{r_{ij} ^2(t+T)} \right) \right\rangle_{i,t},
\label{eqD}
\end{equation}
where the $n_i$ particles $j$ are the neighbors of particle $i$ at time $t$.
Time $T$ is taken to be proportional to the volume of the system, which 
ensures that $\Delta$ records, in the large-size limit, 
an asymptotic property of the system (because $T\gg\tau$).
Clearly, $\Delta\sim 1$ in the ``liquid'' phase for a 
large enough system, while $\Delta\sim 0$ in the 
``solid'' phase. 
Indeed, increasing $\beta$, $\Delta$ falls off sharply (Fig.~\ref{fig02}b).
The transition point is chosen to be at $\Delta=\frac{1}{2}$.

Finally, we must be able to distinguish the regimes for which collective 
motion arises. To this aim, we use the average velocity $V$, as defined
in (\ref{eqV}). 
Increasing $\alpha$, $V$ reaches order-one values. 
The onset of collective motion is chosen to be at $V=v_0/2$. 
In the gas phase, one expects $V=0$ independently of the strengh
of the alignment force. Nevertheless, this phase
is rather sharply divided in two when studying the model at finite
particle density. The largest cluster size may then be small ($n\ll N$), 
but it is (almost) always larger than one.
At any given time, thus, $n>1$, and the collective motion order parameter
$V$ can be defined restricted to the particles belonging to the 
largest cluster. (Since clusters merge and break, the particles involved
generally change along time.)

\section{Phase diagram}

After a brief discussion of the nature of the transitions involved,
we first present the phase diagram of our model for a finite
density of particles in a large and fixed box size.
Then we estimate finite-size
effects on the location of the phase boundaries. Finally, we argue 
that the phase diagram can also be defined in the zero-density limit
where an arbitrarily large flock wanders in an infinite space.

\subsection{Nature of the transitions}

\begin{figure}
\begin{center}
\includegraphics*[width=6.5cm]{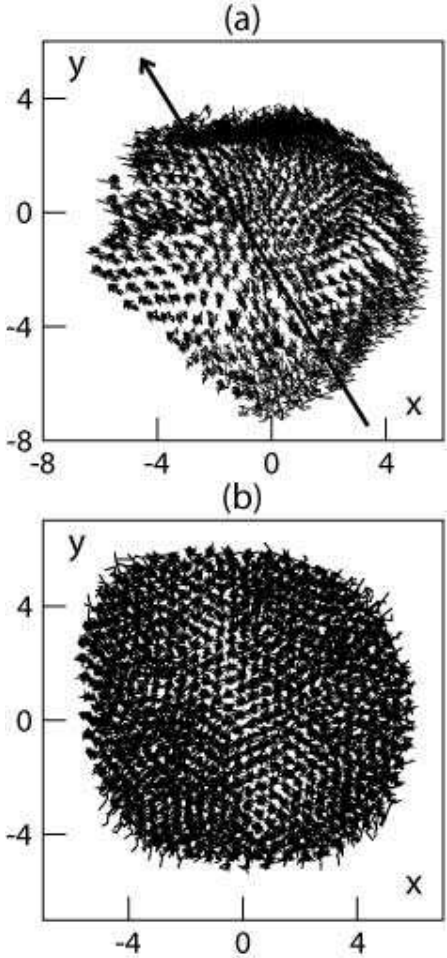}
\caption{Short-time trajectories of freezing droplets of 
$N=512$ boids  ($5000$ timesteps are shown, the motion of the center of mass
and the solid rotation around it have been substracted).
 (a): in the moving phase $\alpha =3.0$, $\beta =70$: the inner part 
appears more solid, while the head is clearly more liquid (the arrow indicates
the instantaneous direction of motion).
 (b): in the non-moving phase $\alpha =1.0$, $\beta =50$: one can distinguish
an outer liquid layer from the almost solid core.}
\label{fig06}
\end{center}
\end{figure}

As expected in usual phase transitions, 
we found that in our model the ``gas/liquid'' and ``liquid/solid'' transitions
are first-order. In insets of Fig.~\ref{fig02}a and b, we show the 
evolution of the probability distribution function of the order parameter as 
one crosses these transition lines.
The bimodal character of these pdf at the transition is typical
of first-order phase transitions, indicating the coexistence of two metastable
states. At the ``gas/liquid'' transition point,
 dispersed and aggregated boids coexist and there are exchanges
between the two phases along time. 
At the ``liquid/solid'' transition point, cohesion is ensured and one observes
the quasi-frozen regions in an otherwise more ``liquid'' flock. The quasi-solid
parts are often located in the core of the flock.
(Fig.~\ref{fig06}).

The nature of the transition for the onset of collective motion
is a delicate issue in our model. Whereas its second-order character
is rather well-established for Vicsek's core model, we recently
discovered that, in fact, the implementation of the noise term
may change the nature of the transition. The detailed 
investigation of this, in particular in presence of the cohesive force,
will be presented in a future publication \cite{TBP}.

\subsection{Fixed population and fixed box size}

A systematic scan of the ($\alpha$, $\beta$) parameter plane was performed
for a flock of $N=2025$ particles living on a square surface of linear size
$L=180$ ($\rho =1/16$) with periodic boundary conditions. Using the criteria
defined above, we obtained the phase diagram presented in Fig.~\ref{fig07}.

For each parameter value, we used an
 initially-aggregated flock.
We let  the system evolve during a time $\tau \propto L^d $, and then we 
recorded each order parameter and its histogram along time. 
The transition points were determined by dichotomy, the precision of
which is reflected in the error bars.

The basic expected features are found: the horizontal ``gas/liquid'' 
and ``liquid/solid'' transitions are crossed by the vertical 
``moving/non-moving'' line. Near this line, however, one observes
a strong deformation of the ``gas/liquid'' and ``liquid/solid'' boundaries.
This cannot be understood without a careful study of the collective motion 
transition \cite{TBP}.

Note also that the ``gas'' phase itself is crossed by the 
line marking the onset of collective motion, using, as explained above,
the average velocity $V$ of the $n$ particles of
the largest cluster as the order parameter.

\subsection{Finite size and saturated vapour effects}

\begin{figure}
\begin{center}
\includegraphics*[width=6.5cm]{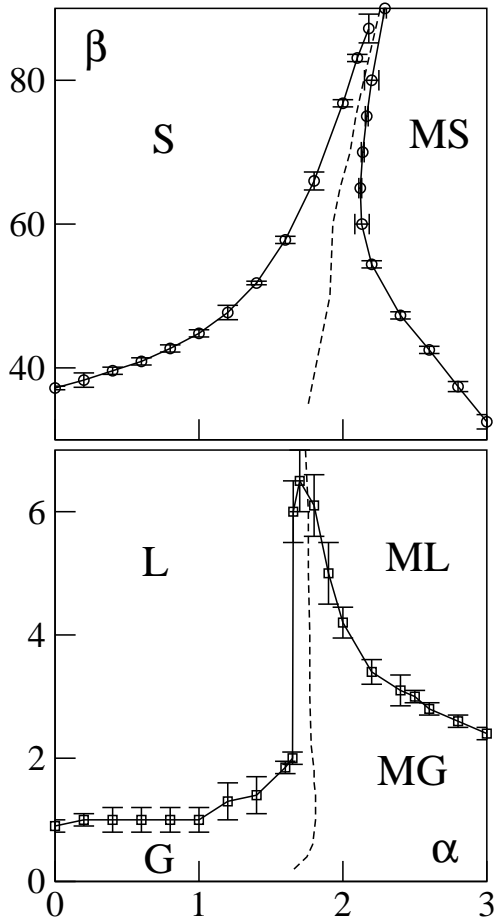}
\caption{Phase diagram at $\rho =1/16$, $L=180$
(other parameters as in Table~\protect\ref{t1}). 
S : solid, MS : moving solid,
L : liquid, ML : moving liquid, G : gas, MG : moving gas. Dashed line : 
transition line of collective motion.}
\label{fig07}
\end{center}
\end{figure}

We are ultimately interested in
the possibility of collective and cohesive motion for an arbitrarily-large
flock in an infinite space. 
The phase diagram of Fig.~\ref{fig07} was obtained at a fixed
system size and constant density. Thus both limits
of infinite-size and zero-density have to be taken
to reach the asymptotic regime of interest. Of course this is mostly relevant
to the onset of cohesion (the ``gas/liquid'' transition).
Here we first study each limit separately, i.e. we investigate 
finite-size effects at fixed particle density and expansion 
at fixed particle number.
Then we discuss the double-limit regime of interest.

Performing such a task for the whole parameter plane far exceeds
our available computer power. We restricted ourselves to three typical
cases:
 in the non-moving phase ($\alpha =1.0$),
 in the moving phase ($\alpha =3.0$) and
 near the transition to collective motion ($\alpha =1.75$ for the 
``gas/liquid'' transition and $\alpha =2.1$ for the ``liquid/solid''
transition). 

In simulations performed at $\rho =1/16$, varying $N$ and $L$,
the transition points $\beta_{\rm GL}^{\rho}(N)$ and  
$\beta_{\rm LS}^{\rho}(N)$ converge exponentially 
as $L=\sqrt{N/\rho}$ increases. In Fig.~\ref{fig08}, we show this for 
both the ``gas/liquid'' (a) and the ``liquid/solid'' (b) 
transitions for $\alpha=3$
(moving phases). This allows to determine asymptotic transition points.
Note that exponentially-decreasing finite-size effects
are typical of first order phase transition at equilibrium \cite{Koct}. 
In the two other cases (non-moving phase,
and onset of motion), the results are similar, with
the asymptotic values being quickly reached  in the non-moving phase 
($\alpha=1$) \cite{these}.

In simulations performed at fixed $N$ varying $L$, we study instead the 
expansion of the system. The (finite) spatial extent induces 
a confinement effect 
which increases the pressure at the coexistence point. We thus expect 
a displacement of the transition point. We find that
$\beta_{\rm GL}(\rho)$ also converges exponentially, and thus
 transition points $\beta_{\rm GL}(N)$ are well-defined.
In Fig.~\ref{fig08}c, we show this only for the ``gas/liquid'' transition 
at the $\alpha$ values used in Fig.~\ref{fig08}a, since
the ``liquid/solid'' transitions points were observed to be independent of
the box size.

\subsection{Zero-density limit}
\label{double-limit}

The above results provide evidence that our model
possesses well-defined, asymptotic phase diagrams at either fixed
particle density or at fixed number of particles.
The double limit mentioned above can be approached in 
essentially three different ways.

\begin{figure*}
\begin{center}
\includegraphics*[width=16cm]{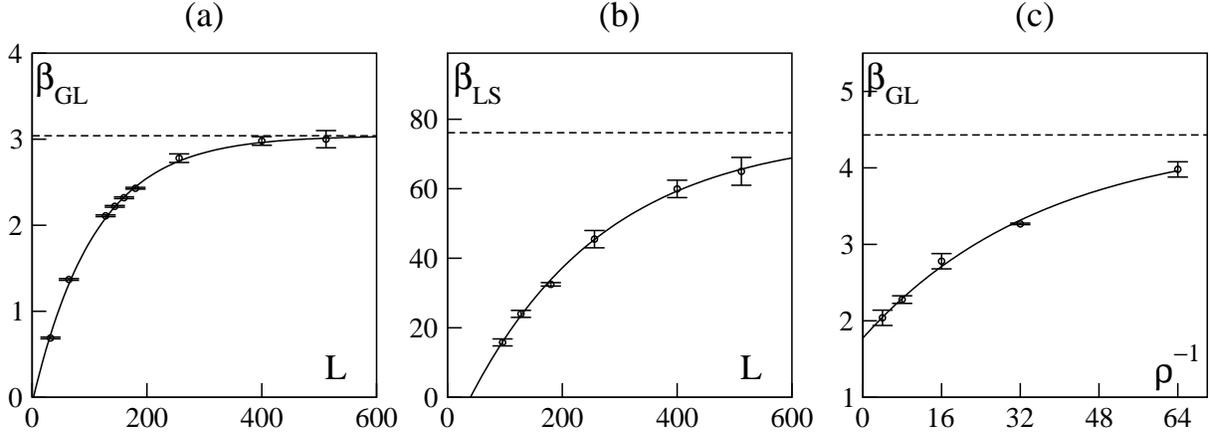}
\caption{Finite size effects on transitions at a fixed density or at fixed 
number of particles at $\alpha =3.0$ (moving phases). 
The dashed lines are the asymptotic values determined by exponential fits
(solid lines).
(a)  variation of the ``gas/liquid'' transition point $\beta_{\rm GL}$
 at $\rho =1/16$ \em vs \em linear size.
(b)  variation of the ``liquid/solid'' transition point
$\beta_{\rm LS}$ at $\rho =1/16$ \em vs \em linear size.
(c) $\beta_{\rm GL}$ vs inverse density $1/\rho$ for $N=4096$.}
\label{fig08}
\end{center}
\end{figure*}

\begin{figure}
\begin{center}
\includegraphics*[width=6.5cm]{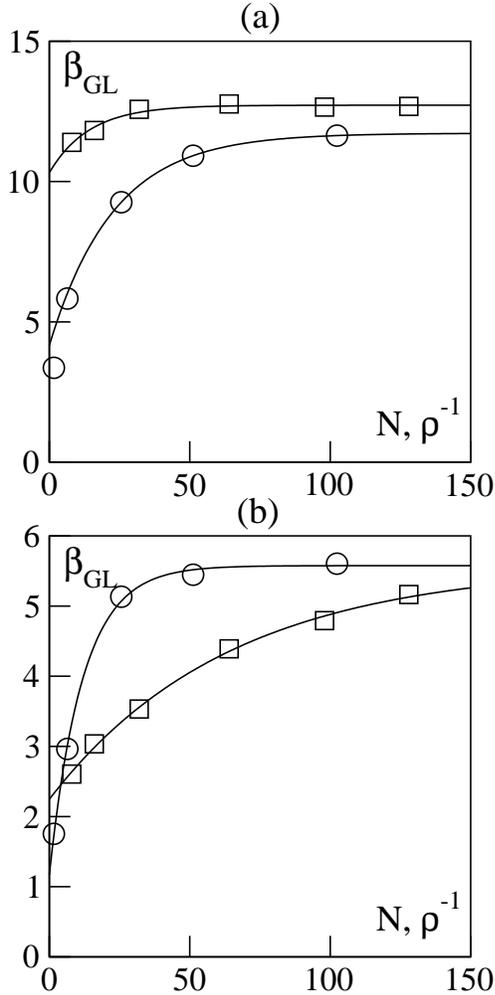}
\caption{Zero-density limit of the cohesion transition 
for (a) $\alpha =1.75$ and 
(b) $\alpha =3.00$ (other parameters as in Table~\protect\ref{t1}). 
Transition points at zero-density for 
different system sizes $N$ (circles), and at infinite-size 
for different (inverse) densities $\rho ^{-1}$ (squares).
The solid lines are exponential fits.
At $\alpha =1.75$, $\beta_{\rm GL}(N) \to 11.7$ and $\beta_{\rm GL}(\rho ) \to 12.7$ ; 
at $\alpha =3.0$, $\beta_{\rm GL}(N {\rm or} \rho ) \to 5.6$.}
\label{fig09}
\end{center}
\end{figure}

One can take one limit after the other one,
repeating either the calculations of Fig.~\ref{fig08}a 
at lower and lower densities,
or those of Fig.~\ref{fig08}d at larger and larger flock size.
However, this straightforward program involves very heavy 
numerical simulations.  
Therefore, we only considered two cases ($\alpha = 1.75$
 and $\alpha =3.0$).
As expected, the transition points converge
(exponentially) independently of the order with which the 
two limits are taken,
yielding estimates of the zero-density limit (Fig.~\ref{fig09}).
These estimates are compatible with each other.
There are three different sources of error: statistical
and systematic errors when evaluating the order parameters of each
system, and then fitting errors when determinating $\beta_{\rm GL}(N)$ 
and $\beta_{\rm GL}(\rho )$. 
For $\alpha =1.75$, we find a $9\%$ difference in the estimated
asymptotic values whose origin is probably the statistical errors on 
the larger flocks simulations.

\subsection{Evaporation of a flock}

There exists a third manner of approaching
the zero-density limit of the ``gas/liquid'' transition.
It consists in quenching a cohesive flock 
observed at large-enough $\beta$ to a lower $\beta$ value.
If the flock is quenched below the ``gas/liquid'' line, it will
``evaporate''.
The largest cluster will progressively lose
particles before finally equilibrating in the gas phase. 
This transient can be expected to be governed by an effective
surface tension $\gamma$ and the boundary of the largest cluster should then
be governed by some Allen-Cahn law \cite{Bray}: $v_n = \gamma \kappa$
where $v_n$ is the (local) normal velocity and $\kappa$ the local curvature.
Assuming that the mass and the surface of the main cluster 
remain proportional, in the
two-dimensional problem: $n=\rho_{\rm loc}\pi R^2 $, we can integrate 
this equation, and
we obtain the relation :
\[
n = N - \lambda t  \;\; {\rm with} \;\; \lambda=\pi \rho _{\rm loc} \gamma
\]
where $\rho _{\rm loc}$ is the local density:
the size of a circular flock should decrease linearly
in time, with the proportionality constant providing an 
estimate of $\lambda$.

\begin{figure*}
\begin{center}
\includegraphics*[width=14cm]{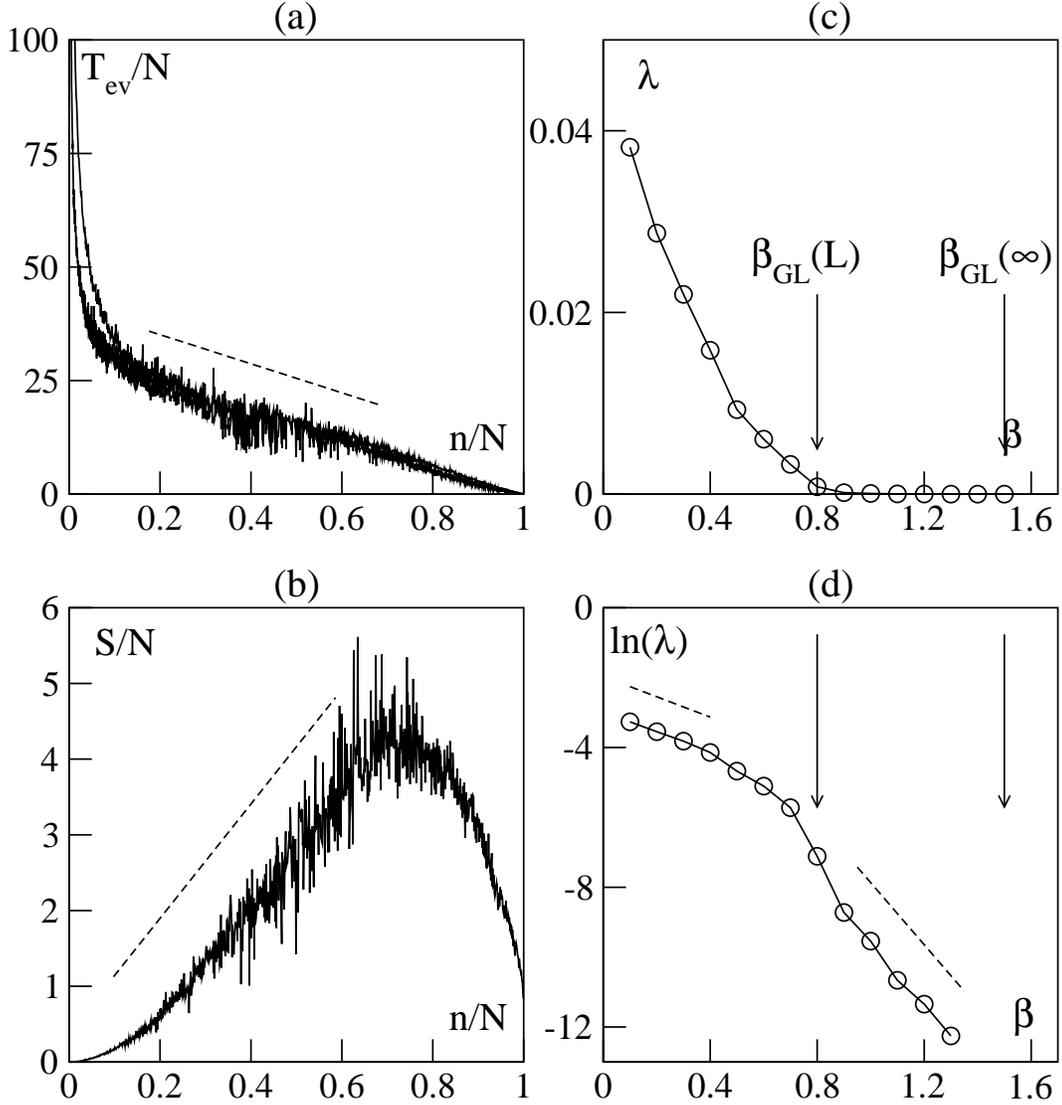}
\caption{Evaporation at $\rho =1/256$, $\alpha =1.0$
(other parameters as in Table~\protect\ref{t1}).
(a) : time \em vs \em normalized cluster mass, $L=512$, and $1024$, 
$\beta =0.2$.
(b) : surface \em vs \em normalized cluster mass, $L=512$, $\beta =0.2$.
(c) surface tension \em vs \em $\beta $, in linear scale, 
(d) same as (c) in log-lin. scales. The dashed lines are only a guide to 
the eye. 
}
\label{fig10}
\end{center}
\end{figure*}

This is indeed what can be observed in our model (Fig.~\ref{fig10}a and 
b).
In these experiments, an initially large cohesive flock is prepared
at some $\beta$ value above the transition. The system size is taken
so as be close to the zero-density limit ($\rho=1/256$
in Fig.~\ref{fig10}).  We measured $T_{\rm ev}$,
the time taken by the largest cluster to
 reach a given normalized mass $n/N$, 
as well as its surface at the same mass/time. 
We thus checked the proportionality between mass and surface 
and between mass and time, after transients and before the system
approaches the equilibrium (Fig.~\ref{fig10}a and b). 
Note that due to the abrupt change of parameters the flock actually 
first expands after the quench before setting in the ``true'' 
evaporation regime (Fig.~\ref{fig10}b).
This experiment can, at first sight, be thought of being
free of confinement effects, and the Allen-Cahn law is expected to
be satisfied at all times.

Figure \ref{fig10}c shows the dependence of $\lambda $ 
on $\beta $: $\lambda$ quickly decays and reaches very small values
for $\beta \sim 0.8$.
On a  logarithmic scale (Fig.~\ref{fig10}d), 
we can distinguish two exponential regimes, on each side of this value,
 which corresponds roughly to the {\it finite-size} threshold 
$\beta_{\rm GL}(L)$ determined above (and is rather far from
the asymptotic threshold determined above to be around 1.5)
A simple argument can account for this behavior:
Consider a flock of $N$ boids and suppose there is no 
short-time expansion (such as seen in Fig.~\ref{fig10}b). 
The mass of the largest cluster $n(t)$, from then on, 
decreases linearly with time until it reaches, at time $T_{\infty}$,
the equilibrium value $n_{\rm eq}$. At every time $t$, we have:
\begin{equation}
N - n(t) = \frac{t}{T_{\infty}} (N-n_{\rm eq}) = \lambda t \;.
\end{equation}
From the theory of first order phase transitions of 
systems at equilibrium (see for instance \cite{Koct}), we 
expect the order parameter $n_{\rm eq}/N$ to behave like
\[ 
\frac{n_{\rm eq}}{N}\sim\frac{1}{2}\left[1+
\tanh \left(K_1 L^d(\beta - \beta_{\rm GL})\right)\right]\,,
\]
 in the vicinity of the transition, where $K_1$ is a constant. 
Moreover, we expect that $T_{\infty}$ depends linearly on $N$
(see Fig.~\ref{fig10}a, where results for two different system sizes
have been superimposed).

From a mean-field point of view, $T_{\infty}/N$ can
be interpreted as the mean time required for a particle to escape 
from the interaction of another boid. Given the interaction we use (see
Eq.~(\ref{bodyforce})), we can assume that the potential is
harmonic, so that the escape time is proportional to the exponential 
of the potential depth $\beta$. Finally, we get
\begin{eqnarray}\nonumber
\lambda&\!\!\!=\!\!\!&\frac{N}{T_\infty}
\left(1-\frac{n_{\rm eq}}{N}\right)\,,\;{\rm or}\\ 
\nonumber
\lambda&\!\!\!\propto\!\!\!&\exp\left(-K_2\beta\right)
\left[1-\tanh\left[K_1 L^d\left(\beta -\beta_{\rm GL}(L) \right)\right]
\right]
\end{eqnarray}
Approximating $\tanh$ by $(-1+\exp)$ and $(1-\exp)$ sufficiently far 
below and above
$\beta=\beta_{\rm GL}(L)$, we find the two exponential regimes mentioned
above. More precisely, for $\beta < \beta_{\rm GL}(L)$, the surface tension
should be almost independent on system size, whereas
for $\beta > \beta_{\rm GL}(L)$, 
$\lambda $ is governed by the finite size effects and its slope
increases like $L^d$. Quantitative agreements with the above
approximation would require too large numerical calculations,
but our partial data is consistent with the above predictions.

\begin{figure}
\begin{center}
\includegraphics*[width=6.5cm]{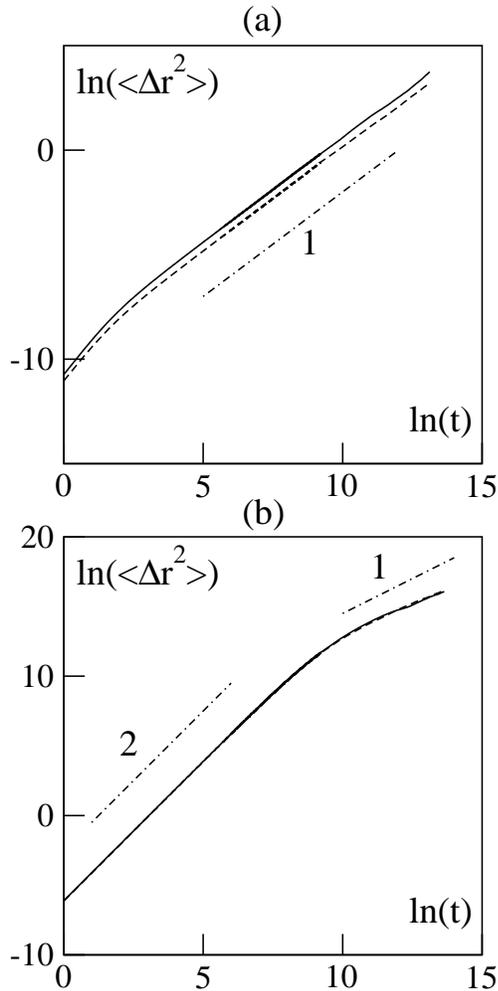}
\caption{Mean-square displacement of the
center of mass vs time for a cohesive droplet (solid lines) and for a
 crystal (dashed lines), $N=32$, $\rho=1/64$, logaritimic scales. 
(a): in the non-moving phase, $\alpha=1$, $\beta=40$ (``liquid'') 
and $65$ (``solid''). 
(b): in the moving phase, $\alpha=3$, $\beta =40$ (``liquid'') and $95$ 
(``solid'').
The straight lines have slope 1 or 2.}
\label{fig11}
\end{center}
\end{figure}
A concluding remark to this investigation of the effective surface tension
governing the evaporation of a flock is that, contrary to naive arguments,
this method does not offer much advantage over the double-limit procedure
presented in Section~\ref{double-limit}. Indeed,
as shown above, it only allows a rather easy determination of 
$\beta_{\rm GL}(L)$, while the finite-size effects remain hard
to estimate quantitatively.

\section{Micro vs macro motion}

The existence of cohesive phases being now well-established,
a natural question is that of the properties of the trajectories
of cohesive flocks (the ``macroscopic'' motion) and it is interesting
to compare those
to the trajectories of the individuals composing the flock
(``microscopic'' motion). Postponing again the account of what happens
in this respect near the onset of collective motion to a further 
publication \cite{TBP}, we studied, for the four possible cohesive 
phases, the mean square displacement $\langle\Delta r^2(t)\rangle$
 of the center of mass, as well as of individual boids. 

\begin{figure*}
\begin{center}
\includegraphics*[width=14cm]{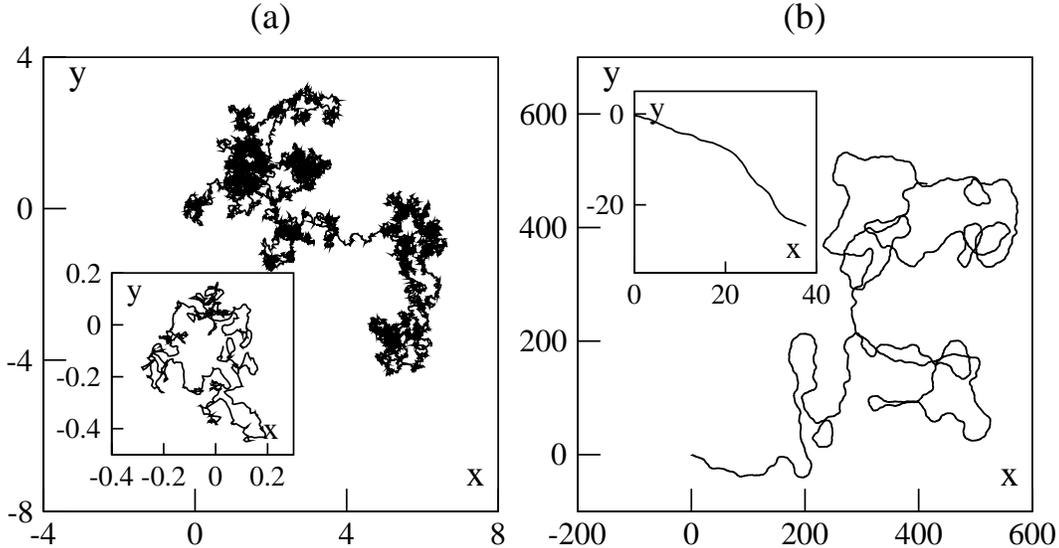}
\caption{Trajectories of the center of mass in the non-moving phase
((a) : $\alpha =1$) and in the moving phase ((b) : $\alpha =3$).
Flock of $N=32$ boids, $\rho =1/256$, $\beta =40.0$.
At long times (a,b) $10^5$ timesteps. 
At short times (insets): 1000 timesteps shown.
}
\label{fig12}
\end{center}
\end{figure*}

\begin{figure}
\begin{center}
\includegraphics*[width=5cm]{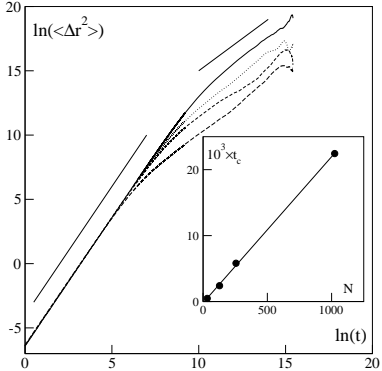}
\caption{Mean square displacement of the center of mass vs time 
for a moving droplet
of size $N= 1024$, $256$, $128$, and $32$ from top to bottom 
(logarithmic scales,  $\rho =1/64$, $\alpha=1.95$, $\beta=40$).
The solid lines have slope 2 and 1. The larger the flock the more ballistic
its motion at short times. The inset represents the cross--over 
time between the ballistic and the brownian motion.}
\label{fig13}
\end{center}
\end{figure}

Our model being essentially stochastic, it is no surprise that,
at large times, we observe that  $\langle\Delta r^2(t)\rangle \sim t$, 
i.e. the flock performs Brownian motion, in all cases (Fig.~\ref{fig11}
 and \ref{fig12}). 
When in a moving phase (either ``liquid'' or ``solid''),
this random walk may consist of ballistic flights separated by 
less coherent intervals during which the flock often changes direction.
This is testified by the ballistic part 
($\langle\Delta r^2(t)\rangle \sim t^2$)
of the plots in Fig.~\ref{fig11}b and by the trajectories itselves 
\ref{fig12}. 
The moving and non-moving phases
can also be distinguished by opposite finite-size effects: in the non-moving
phases, the diffusion constant of the macroscopic random walk
decreases with system size (like $1/N$), whereas the ballistic
flights' duration increases with system size in the moving phases
(Fig.~\ref{fig13}).

Comparing the macroscopic motion of ``liquid'' and ``solid'' flocks 
is not well defined since, in the phase diagram, the line marking
the onset of collective motion is not straight. Nevertheless, 
at comparable distances from this line, one notices that
ballistic flights tend to be longer for flying crystals than for 
moving droplets. Similarly, the diffusion constant of ``solid'', non-moving
flocks is smaller than that of non-moving droplets.

\begin{figure}
\begin{center}
\includegraphics*[width=4.8cm]{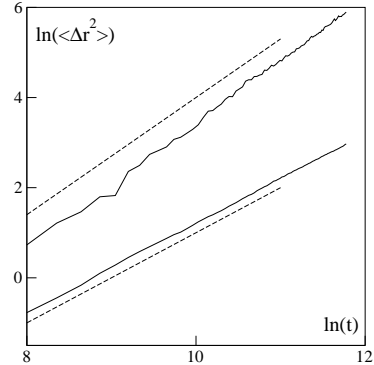}
\caption{Internal mean-square displacement vs time of the individual boids 
of a cohesive droplet of 10000 boids ($\rho=1/16$, $\beta=40$,
the motion of the center of mass has been substracted).
Lower curve: non-moving case ($\alpha=1$), 
normal diffusion (the associated dashed line has slope 1).
Top curve: moving case ($\alpha=3$), 
superdiffusion (the associated dashed line has slope 4/3, 
as predicted by Toner and Tu \protect\cite{TT2}).
}
\label{fig14}
\end{center}
\end{figure}

Finally, even though we have already considered the mutual dispersion
of initially-neighboring boids (see Fig.~\ref{fig03}), we also
studied the diffusive properties of individual boids within cohesive flocks,
substracting out the translation motion
of the center of mass. In ``solid'' flocks, one can hardly record any
such motion. For droplets (Fig.~\ref{fig14}), on the other hand,
this microscopic motion depends on the macroscopic motion. When the 
droplet is fixed, the diffusion is normal $\langle\Delta r^2\rangle\sim t$, 
whereas a boid which belongs to moving flock diffuses as 
 $\langle\Delta r^2\rangle\sim t^{\alpha}$, with $\alpha \sim 4/3$.
We interpret this as being due to some mesoscopic ``hydrodynamical''
structures within moving flocks (jets, vortices, etc.). J. Toner and Y. Tu
\cite{TT2} have shown \em via \em a mesoscopic equation 
that collective motion induces transverse correlations 
even in a co-moving frame. Therefore, boid 
diffusion must be faster than Brownian. They predicted an exponent equal 
to $4/3$, with which our results are in good agreement (Fig.~\ref{fig14},
top line).

\section{Summary and Perspectives}

We have introduced a simple model for the collective and cohesive
motion of self-propelled particles. We have described its various
dynamical phases, defined order parameters to distinguish them,
and presented a typical phase diagram at large but finite number of particles
$N$ and large but finite system size $L$ (Fig.~\ref{fig07}). 
Even though we have provided evidence that this phase diagram 
possesses well-defined $N\to\infty$ and $L\to\infty$ limits, these
limit diagrams require too heavy numerical simulations to be determined
at this stage. 

We have also argued that the double limit of an arbitrarily large flock
evolving in an infinite space is also well-defined. Here also,
the mapping of the phase diagram in this limit is currently out of reach.
But the existence of cohesive phases in a model of noisy, short-range
interaction, identical particles is ensured, which was one of our 
primary goals. Futhermore, we can sketch the zero-density asymptotic
diagram from our partial knowledge (Fig.~\ref{fig15}).
A few remarks are in order to explain the expected shape of the 
asymptotic ``gas/liquid'' boundary: in the non-moving
phase, this transition is almost independent on $\alpha$ and 
size-effects are negligible. We thus expect this line to be horizontal,
in agreement with mean-field arguments (see Appendix).
Similarly, our preliminary study of the onset of collective motion shows
that one can go directly from a cohesive non-moving droplet to the incohesive
phase and that in this region the onset of collective motion is 
roughly independent on $\beta$, yielding a vertical boundary.

More work should be devoted to the determination of the 
asymptotic phase diagrams mentioned above, as well as to
a quantitative study of the onset of collective motion.  Even in the
simplest case of the non-cohesive Vicsek-type models, it can be 
second or first order, depending on the nature of the microscopic 
noise in the model \cite{TBP}.
In both cases, we have 
started to uncover a rich interplay between collective motion,
critical fluctuations, rotation modes, and shape dynamics in the 
transition region.

Not even mentioning the study of our model in three dimensions and its
possible applications to particular real-world situations (e.g. biology,
zoology, and robotics) the work 
presented here is probably only the beginning of the exploration of
this new type of non-equilibrium systems.

\section*{Appendix: Mean-field approach} 

To understand the interplay between alignment and body force, we
 simplify our system \em via \em a Hamiltonian model. The 
total energy is the sum of kinetic energy, two-body interaction energy, 
and a term related to the velocity alignment ``force''. We thus write 
\begin{eqnarray}\nonumber
{\cal H}=\frac{Nm}{2}v_0^2&\!\!\!\!+\!\!\!\!&U\lpa r_1,r_2,...,r_N\rpa
\\
&\!\!\!\!-\!\!\!\!&\alpha\sum_{i,j\sim i}^N\vec{v}_i.\vec{v}_j
-\vec{h}_0.\sum_i^N\vec{v}_i\,.\nonumber
\end{eqnarray}
where we have introduced an external field $\vec{h}_0$.

\begin{figure}
\begin{center}
\includegraphics*[width=6.5cm]{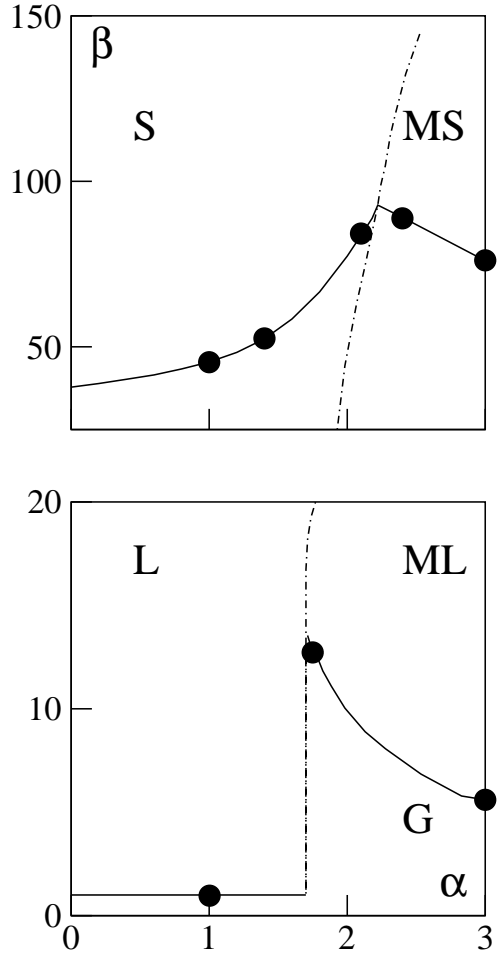}
\caption{Sketch of the asymptotic phase diagram in the zero density limit.
Filled circles indicate points determined numerically as in 
Section~\protect\ref{double-limit}. See text for more details.}
\label{fig15}
\end{center}
\end{figure}

In the spirit of the mean-field approach, 
we assume that the fluctuations are negligible so that each component
 can be integrated independently and approximated  by its 
averaged value. Thus the contibution from the two-body interaction 
becomes:
\begin{eqnarray}\nonumber
U\lpa r_1,...,r_N\rpa&\!\!\!\sim\!\!\!&\frac{1}{2}
\sum_{i\sim j}u\lpa|\vec{r}_i-\vec{r}_j|\rpa\sim\frac{N}{2}U_0\,,\\
U_0&\!\!\!=\!\!\!&\int_{R_{\rm c}}^{+\infty}\beta\rho u(r)d^dr
\sim-2a\rho\beta\,,\nonumber
\end{eqnarray}
where $a$ is a constant which depends on the potential. 
The alignment 
energy is computed with the coarse--grained velocity
$\vec{\varphi}=\lbt\vec{v}\rbt$ and the average number of neighbors 
$\rho\pi r_0^2=\rho s$ :
\[
\alpha\sum_{i,j\sim i}^N\vec{v}_i.\vec{v}_j
+\vec{h}_0.\sum_i^N\vec{v}_i\sim\lpa\vec{h}_0+\alpha\rho s\vec{\varphi}\rpa
.\sum_i^N\vec{v}_i\,.
\]
In the following, we use $\vec{h}_0+\alpha\rho s\vec{\varphi}=\vec{h}$.

Because of the separation of
phase space variables, each term of the Hamiltonian contributes
a factor to the partition function:\\
\noindent
$\bullet$ the two-body coupling :
\bequ
\lpa S-bN\rpa^N\exp\lpa aN\rho\frac{\beta}{kT}\rpa\,,
\eequ
where $b=\frac{\pi}{2}R_c^2$ is the hard--core surface and $S$ is the 
whole surface,\\
\noindent
$\bullet$ the alignment coupling :
\bequ
\lcr 2\pi mv_0{\rm I_0}\lpa -\frac{hv_0}{kT}\rpa\rcr^N\,,
\eequ
where ${\rm I}_0$ is a Bessel function.
The free energy of the system is thus written:
\barr\nonumber
-\frac{F}{NkT}\!\!\!\!&=&\!\!\!\!\ln\lpa\frac{mv_0}{\hbar}\rpa
-\frac{mv_0^2}{2kT}+1
+\ln\lpa \rho^{-1}-b\rpa\\
\!\!\!\!&+\!\!\!\!&a\rho\frac{\beta}{kT}+\ln{\rm I}_0\lpa 
-\frac{hv_0}{kT}\rpa-\rho s\varphi^2\frac{\alpha}{2kT}
\nonumber
\earr
Imposing to be at a minimum of the free energy,
we find that the  average velocity is the solution of a 
self--consistent equation: 
\bequ
\frac{\varphi}{v_0}=\frac{{\rm I}_1}{{\rm I}_0}\lpa \frac{v_0h}{kT}\rpa\,,
\eequ
and that the critical point is defined by
\bequ
sv_0^2\rho\alpha=2kT\,.
\eequ
Collective motion thus emerges via a continous phase transition 
whose critical exponents are $1/2$ for the order parameter and $1$ 
for its susceptibility, independently of the cohesive force.

\begin{figure}
\begin{center}
\includegraphics*[width=6cm]{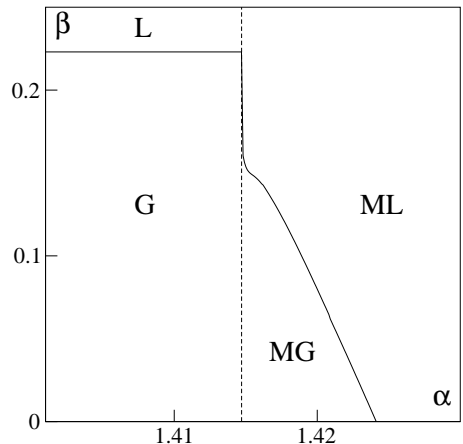}
\caption{Phase diagram of a mean field model.
L : liquid, ML : moving liquid, G : gas, MG : moving gas. Dashed line : 
transition line of collective motion.}
\label{fig16}
\end{center}
\end{figure}

Furthermore, the necessary concavity of the free energy function 
implies a first order cohesion transition, 
since the second derivative of the free energy 
has two zeros. Their positions define the phase coexistence region.
We determined the stability limit of the gas phase. 
As positions and velocities 
are decorrelated, we studied two cases : first without any motion, and then 
with collective motion. In the first case:
\bequ
\beta_{\rm GL}=\frac{kT}{2a\rho\lpa 1-b\rho\rpa^2}\,,
\eequ
which means that the transition line does not depend on the alignment 
parameter (Fig.~\ref{fig16}). Note also that there is no transition point 
in the zero--density limit of this model. We numerically solved
the case with collective motion and found a stabilization of the ``liquid''
 phase (Fig.~\ref{fig16}). 
This is an effect of the assumption that velocity fluctuations are negligible.
At the onset of motion, we expect that fluctuations of the macroscopic
velocity diverge, leading to strong density fluctuations. This 
could explain the de--stabilization of the ``liquid'' phase (Fig.~\ref{fig15}) 
and its stabilization in the mean field model~\ref{fig16}, 
where such fluctuations are by definition absent.


\begin{thebibliography}{10}

\bibitem{Parrish} J.K. Parrish and W.M. Hamner (Eds.), \em Three dimensional animals 
groups \em, Cambridge University Press (1997), and references therein.


\bibitem{VICSEK} T. Vicsek, A. Czir\'ok, E. Ben-Jacob, I. Cohen and O. 
Shochet,
 Phys. Rev. Lett. {\bf 75}, 1226 (1995).
A. Czir\'ok, H. E. Stanley, T. Vicsek, J. Phys. A {\bf 30},
1375 (1997).

\bibitem{Reynolds} C. W. Reynolds, Comput. Graph. {\bf 21}, 25 (1998).

\bibitem{Shimoyama} N. Shimoyama, K. Sugawara \em et al \em, Phys. Rev. Lett {\bf 76}, 3870 (1996).

\bibitem{Levine}
A. S. Mikhailov and D. H. Zanette,  Phys. Rev. E {\bf 60},
4571 (1999);
H. Levine, W.-J. Rappel and I Cohen, Phys. Rev. E {\bf 63}, 017101 (2001);
I. Couzin \em et al \em, J. theor. Biol {\bf 218}, 1 (2002). 

\bibitem{Duparc} Y. L. Duparcmeur, H. Herrman and J. P. Troadec, 
 J. Phys. I France {\bf 5}, 1119 (1995); 
J. Hemmingson, J. Phys. A {\bf 28}, 4245 (1995).


\bibitem{TT} J. Toner and Y. Tu, Phys. Rev. Lett. {\bf 75}, 4326 (1995);
Phys. Rev. E {\bf 58}, 4828 (1998)

\bibitem{TT2} J. Toner, Y. Tu and Ulm, Phys. Rev. Lett. {\bf 80}, 4819 (1998)

\bibitem{WL} X.-L. Wu and A. Libchaber, Phys. Rev. Lett. {\bf 84}, 3017 
(2000).

\bibitem{COMMENT} G. Gr\'egoire, H. Chat\'e, and Y. Tu,
Phys. Rev. Lett. {\bf 86}, 556 (2001) and Phys. Rev. E {\bf 64}, 011902 (2001).

\bibitem{WL-REPLY} X.-L. Wu and A. Libchaber, Phys. Rev. Lett. {\bf 86},
 557 (2001)

\bibitem{TBP} G. Gr\'egoire, H. Chat\'e, and Y. Tu, in preparation.

\bibitem{these}G. Gr\'egoire, \em Mouvement collectif et physique hors 
d'\'equilibre\em, PhD thesis, Universit\'e Denis Diderot--Paris 7, Paris 
(2002).

\bibitem{Voronoi} A. Okabe \em et al\em, \em Spatial Tesselations, ed. 
J. Wiley and sons\em, ltd, Chichester (1992).

\bibitem{HK} J. Hoshen and R. Kopelman, Phys. Rev. B {\bf 14}, 3438 (1976)

\bibitem{Koct} C. Borgs and R. Koteck\`y, J. Stat. Phys. {\bf 61}, 79 (1990)

\bibitem{Bray} See, e.g.: A.J. Bray, Adv. Phys. {\bf 43},357 (1994)

\end{thebibliography}
\end{document}